\documentclass[%
 reprint,%
 amssymb, amsmath,%
 aip,cha,%
]{revtex4-1}
\usepackage{graphicx}
\usepackage{dcolumn}
\usepackage{bm}
\usepackage{docs}%
\usepackage{bm}%
\usepackage{subfigure}
\usepackage[colorlinks=true,linkcolor=blue]{hyperref}%
\expandafter\ifx\csname package@font\endcsname\relax\else
 \expandafter\expandafter
 \expandafter\usepackage
 \expandafter\expandafter
 \expandafter{\csname package@font\endcsname}%
\fi
\begin{document}

\title{Confining rigid balls by mimicking quadrupole ion trapping}%

\author{Wenkai Fan}
\affiliation{Kuang Yaming Honors School, Nanjing University}

\author{Li Du\footnote{Current Address: Department of Physics, Massachusetts Institute of Technology}}
\email{lidu@mit.edu}
\affiliation{School of Physics, Nanjing University}

\author{Sihui Wang}
\email{wangsihui@nju.edu.cn}
\affiliation{School of Physics, Nanjing University}

\author{Huijun Zhou}%
\affiliation{School of Physics, Nanjing University}

\date{\today}%
\revised{-}%

\begin{abstract}

The rotating saddle not only is an interesting system that is able to trap a ball near its saddle point, but can also intuitively illustrate the operating principles of quadrupole ion traps in modern physics. Unlike the conventional models based on the mass--point approximation, we study the stability of a ball in a rotating--saddle trap using rigid--body dynamics. The stabilization condition of the system is theoretically derived and subsequently verified by experiments. The results are compared with the previous mass--point model, giving large discrepancy as the curvature of the ball is comparable to that of the saddle. We also point out that the spin angular velocity of the ball is analogous to the cyclotron frequency of ions in an external magnetic field utilized in many prevailing ion--trapping schemes.

\end{abstract}

\maketitle


\section{Introduction}

The rotating saddle has a number of intriguing mechanical properties such as the counterintuitive stabilization of particles in the vicinity of its saddle point,\cite{brouwer1918motion,rueckner1995rotating, thompson2002rotating, bottema1976stability} the unexpected precession due to a Coriolis-like force in the inertial reference frame\cite{kirillov2016rotating} and so on. What endows this demonstration with an even profounder meaning is the application of its underlying physical principle to the field of ion trapping. As illustrated by Wolfgang Paul in his Nobel Lecture,\cite{paul1990electromagnetic} by rotating or vibrating a ``saddle--like'' electrostatic potential (a.k.a. the quadrupole potential), one can realize a stable equilibrium and thus confine ions in a vacuum chamber. Such analogy between mechanics and electromagnetics ingeniously interprets the trapping mechanism of the Paul trap---a prototype of the quadrupole ion trap family---in an intuitive way, which touches the frontier of many areas among atomic physics,\cite{amoretti2002production, foot2004atomic, savard1991new, rempe1987observation, sauter1986observation} plasma physics,\cite{itano1998bragg, hornekaer2002formation, dubin1999trapped} quantum computation\cite{cirac1995quantum, kielpinski2002architecture, duan2001geometric} and so on.

\begin{figure}[h]
\centering
\includegraphics[width=0.5 \textwidth]{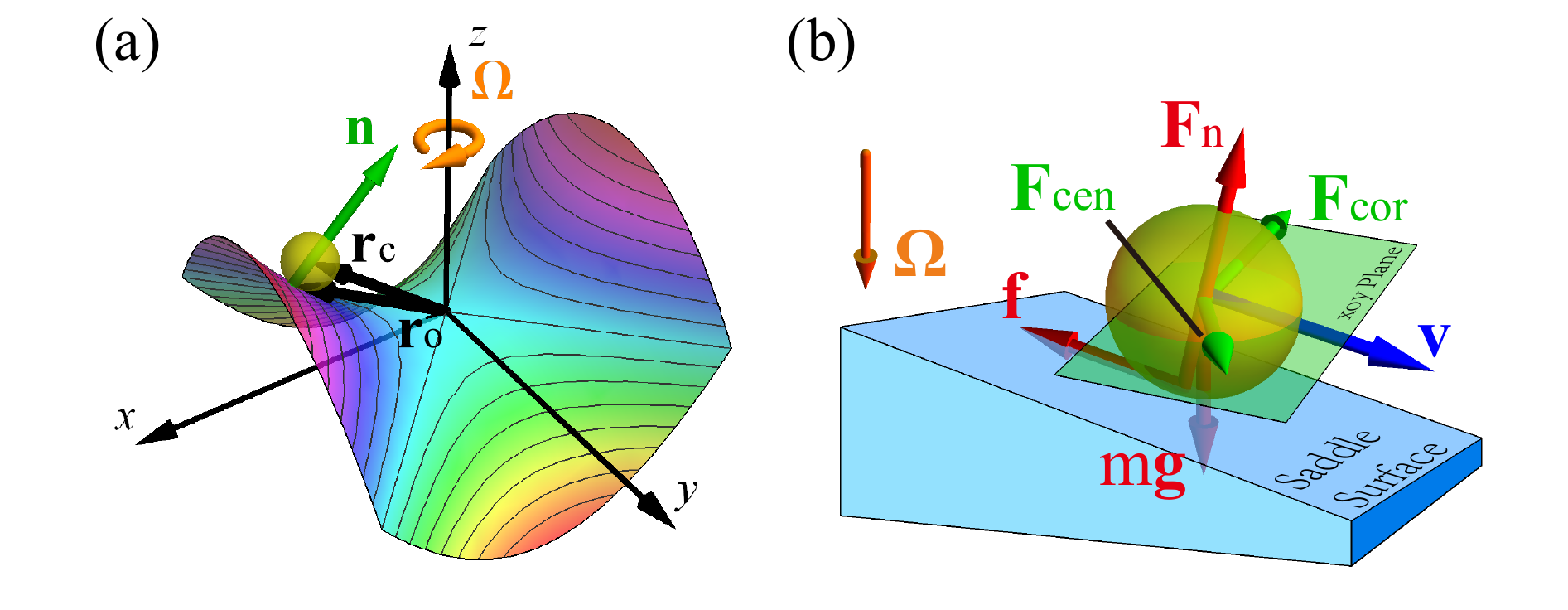}
\caption{(a) Geometric illustration of the rotating-saddle trap. (b) Force diagram in the rotating frame. The gravitational force $m\mathbf{g}$, the supporting force $\mathbf{F_{n}}$, the frictional force $\mathbf{f}$ acted by the saddle, the centrifugal force $\mathbf{F_{cen}}$ and the Coriolis force $\mathbf{F_{cor}}$ arising from the rotation of the reference frame, are exerted on the ball.}
\label{fig:forceDiagram}
\end{figure}

The stability of particles on a rotating surface was studied and reviewed by several early theoretical papers in detail.\cite{brouwer1918motion, bottema1976stability} As a special case, the full set of conditions for stabilization was further applied to the saddle--shaped surface by L.E.J. Brouwer in his 1918 pioneering work.\cite{brouwer1918motion} The surface equation of a saddle one typically considers is\footnote{The general saddle--shaped surface equation should be $z=x^{2}/a-y^{2}/b$ where $a$ and $b$ are positive constants. For the sake of simplicity, however, we consider a special case where $a=b$ (symmetric saddle) in this article without losing generality.}
\begin{equation}
F(x,y,z)=z-\frac{1}{a}(x^2-y^2)=0
\label{eq:SaddleSurface}
\end{equation}
where $a$ ($>0$), $x$, $y$ and $z$ all possess the dimension of length. Apparently, the saddle point of the surface locating at $x=y=0$ is an unstable equilibrium in the static case (see Fig.~\ref{fig:forceDiagram}(a)). But as firstly derived by Brouwer\cite{brouwer1918motion} and subsequently discussed by many other papers,\cite{bottema1976stability, thompson2002rotating} a mass--point constrained on this saddle can be stabilized (without slipping away from the equilibrium point) when the angular velocity $\Omega$ of the saddle exceeds a critical value $\Omega_{crt}$:
\begin{equation}
\Omega \geq \Omega_{crt} = \sqrt{\frac{2g}{a}}
\label{eq:MassPointThreshold}
\end{equation}
where $g \approx 9.8 m/s^{2}$ is the gravitational acceleration.

Albeit intuitive and inspiring, it can be easily noticed that the mass--point model exhibits some important deficiencies when accounting for many experimental phenomena. First of all, it was demonstrated in early experiments\cite{rueckner1995rotating} that both the radius of the ball $R$ and the curvature of the saddle $2/a$ can significantly influence the stability of the system. But the stabilization condition Eq.~\ref{eq:MassPointThreshold}, in which the radius $R$ is apparently absent, cannot describe the size effect of the balls. Secondly, most previous works treated the motion of the ball as a two--dimensional problem, and therefore the dynamic constraint for the rigid ball to stay on the saddle surface was neglected. As a consequence, the high--speed instability---a phenomenon frequently observed where the ball jumps off the surface as the saddle rotates fast enough---has not yet been well--explained.

In this paper, we establish a rigid--body model for the rotating--saddle problem starting from deriving the rigorous equation of motion of the ball on a rotating surface with arbitrary shape. We linearize the equations in the vicinity of the equilibrium point to obtain the stabilization condition. The resulting lower limit of the rotating speed for the onset of trapping is found congruent with that of the mass--point model as the radius $R\rightarrow 0$. By investigating the interacting force between the saddle and the ball, we explain why the ball tends to jump off the saddle when the rotating speed is high. Finally, by comparing the rotating--saddle trap with several quadrupole ion--trapping schemes, we figure out an appropriate electrical analogy to our rigid body model.

The rotating saddle trap provides a fantastic teaching example in undergraduate classes. It illustrates how stabilization can be achieved by rotating a saddle--like potential, either mechanically or electrically, that would otherwise be unstable in static case. The model may simply be demonstrated using the mass--point approximation, which can lead to a basic understanding of the trapping mechanisms. However, the rigid--body model provides a more accurate description of its dynamics, as well as a more challenging problem. The analogy between the mechanical trap and ion traps brings connection between classroom demonstration and the modern techniques in capturing charged particles widely used in physics research today.

\section{Mechanical Model}

We consider a rigid ball with radius $R$ rolling on a saddle surface rotating with angular velocity $\bm{\Omega}$. The coordinate system is chosen to be fixed on the saddle as illustrated in Fig.~\ref{fig:forceDiagram}(a), thus we are able to avoid many time--dependent terms but, as a consequence, acquire the centrifugal and the Coriolis forces in such a non--inertial frame.

\begin{figure}[htbp]
\centering
\includegraphics[width=0.28 \textwidth]{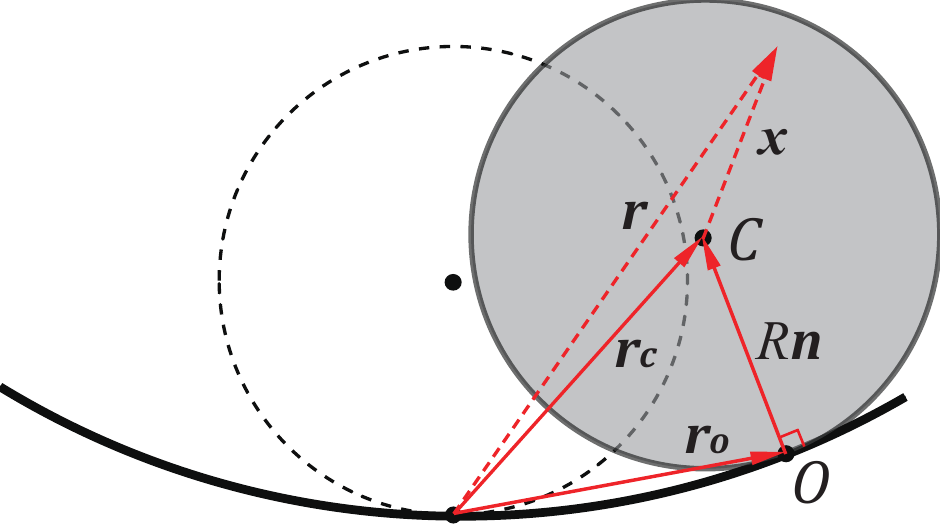}
\caption{Geometric illustration of position vectors. The dashed circle represents the equilibrium position of the ball.}
\label{fig:posVectors}
\end{figure}

The motion of the rigid ball is described by its spatial position $\bm{r}=x\bm{i}+y\bm{j}+z\bm{k}$ and spin angular velocity $\bm{\omega} = \omega_{x}\bm{i}+\omega_{y}\bm{j}+\omega_{z}\bm{k}$ ($\bm{i}$, $\bm{j}$ and $\bm{k}$ are unit vectors along $x$, $y$ and $z$ directions). The total degrees of freedom of the system are reduced due to the constraints provided by the saddle surface. One of the constraints comes from the requirement that the ball stays on the surface of the saddle, and therefore the position vectors of the contact point $O$ and its center of mass $C$ have to satisfy
\begin{equation}
\bm{r_{c}}=\bm{r_{o}}+R\bm{n}
\label{eq:PositionConstraion}
\end{equation}
where $\bm{r_{c}}$, $\bm{r_{o}}$ and $\bm{n}$ denote the position vectors of the mass center of the ball, the contact point with the saddle surface, and the unit normal vector at the contact point, respectively (see Fig.~\ref{fig:posVectors}). The explicit form of the normal vector in Cartesian coordinates can be calculated as
\begin{equation}
\bm{n}=\frac{\nabla F}{|\nabla F|}=\frac{1}{\sqrt{4x_{o}^{2}+4y_{o}^{2}+a^{2}}}(-2x_{o}\bm{i}+2y_{o}\bm{j}+a\bm{k})
\label{eq:NormalVector}
\end{equation}
where $F(x,y,z)$ is the surface equation described in Eq.~\ref{eq:SaddleSurface}.

Another constraint comes from the interaction between the orbital motion of the mass center of the ball and its spin---we call it the spin--orbit interaction. Since the real motion of the ball is a combination of slipping and rolling, we consider two limiting cases: the pure--rolling and the frictionless--slipping cases, which correspond to the strongest and the weakest spin--orbit interactions. If the ball is rolling on the surface without slipping, then the pure--rolling condition yields the following relation between its orbital velocity $d\bm{r_{c}}/dt$ and its spin $\bm{\omega}$
\begin{equation}
\frac{d\bm{r_{c}}}{dt}=\bm{\omega}\times R \bm{n}
\label{eq:PureRolling}
\end{equation}
On the other hand, if the motion of the ball is of pure frictionless--slipping, the angular velocity of the ball is an ignorable coordinate (which remains unchanged over time) that is only determined by its initial condition
\begin{equation}
\bm{\omega}=\bm{\omega_{0}}
\label{eq:PureSlipping}
\end{equation}

With the geometric and kinematic constraints being clarified, the torque equation can be derived with respect to the contact point $O$, having the virtue that the torques provided by the supporting and the frictional forces are vanished (see Fig.~\ref{fig:forceDiagram}(b)). But one should be very careful that, since the contact point $O$ is not a fixed point, the commonly--used torque equation $\bm{M_{o}}=d\bm{L_{o}}/dt$ is no longer valid for our system. Instead, a modification term should be added as follows (see Appendix.~\ref{appendix:A} for derivation):
\begin{equation}
\bm{M_{o}}=\frac{d \bm{L_{o}}}{dt}+\frac{d \bm{r_{o}}}{dt}\times m \frac{d \bm{r_{c}}}{dt}
\label{eq:TorqueAM}
\end{equation}
where the left--hand--side is the torque about point $O$, and the right--hand--side is composed of two terms including (i) the rate of change of the angular velocity about point $O$, and (ii) the rate of change of the angular velocity due to the time--dependence ($d\bm{r_{o}}/dt\neq 0$) of the reference point $O$.

The explicit form of the angular momentum about point $O$ in Eq.~\ref{eq:TorqueAM} is
\begin{equation}
\bm{L_{o}}=mR\bm{n}\times \frac{d\bm{r_{c}}}{dt}+I\bm{\omega}
\label{eq:AM}
\end{equation}
where $I=\alpha m R^{2}$ ($\alpha = 2/5$ for solid spheres) is the moment of inertia of the ball; the first term on the right--hand--side is the contribution from the orbital motion, and the second term arises from the spin of the ball about its mass center. Since the ball is constrained on the saddle surface, we can further know that the velocity of the contact point $O$ is always parallel to the velocity of the center of mass $C$, and therefore $d \bm{r_{o}}/dt\times d \bm{r_{c}}/dt=0$ in Eq.~\ref{eq:TorqueAM}.

We now calculate the torque on the ball with respect to the contacting point $O$. The torque provided by gravitational force is simply $\int_{V}(\textbf{r}-\textbf{r}_{o})\times \bm{g} dm=mR\textbf{n}\times \bm{g}$, as in the inertial frame. Due to the position and velocity dependence of the inertia forces (centrifugal force and Coriolis force), they have an inhomogeneous distribution on the ball. For an infinitesimal segment with mass $dm$ at position $\bm{r}$, the torques provided by centrifugal force and Coriolis force are:
\begin{equation}
\left\{
\begin{aligned}
&d\bm{M_{cen}}=-(\bm{r}-\bm{r_{o}})\times (\bm{\Omega} \times(\bm{\Omega}\times \bm{r}))dm\\
&d\bm{M_{cor}}=-2(\bm{r}-\bm{r_{o}})\times (\bm{\Omega}\times\frac{d\bm{r}}{dt})dm
\end{aligned}
\right.
\label{eq:TorqueSegment}
\end{equation}
To make better use of the symmetry of the ball, we introduce the relative position $\bm{x}$ from a mass segment (located at $\bm{r}$) to the spherical center (located at $\bm{r_{c}}$): $\bm{x}=\bm{r}-\bm{r_{o}}-R\bm{n}$ (see Fig.~\ref{fig:posVectors}). Thus it is not difficult to evaluate the body integral of Eq.~\ref{eq:TorqueSegment} by integrating over $\bm{x}$, which yields:
\begin{equation}
\left\{
\begin{aligned}
&\bm{M_{cen}}=\int_{V}d\bm{M_{cen}}=-mR\bm{n}\times (\bm{\Omega} \times (\bm{\Omega} \times \bm{r_{c}}))\\
&\bm{M_{cor}}=\int_{V}d\bm{M_{cor}}=-2mR\bm{n}\times (\bm{\Omega}\times\frac{d\bm{r_{c}}}{dt})+I\bm{\omega}\times\bm{\Omega}
\end{aligned}
\right.
\label{eq:Torques}
\end{equation}

Notably, for a rigid body possessing spherical symmetry, the expression of $\bm{M_{cen}}$ is the same as that acting on a mass--point; however, an additional term $I\bm{\omega}\times\bm{\Omega}$ appears in the expression of $\bm{M_{cor}}$. This extra term is due to the asymmetrical distribution of the Coriolis force over the ball, which distinguishes the rigid--body model from the conventional mass--point model.

By substituting Eq.~\ref{eq:AM} and Eq.~\ref{eq:Torques} into Eq.~\ref{eq:TorqueAM}, we finally get the equation of motion for the rigid ball moving on an arbitrary surface in the rotating frame:
\begin{equation}
\begin{aligned}
\frac{d}{dt}(mR\bm{n}\frac{d\bm{r_{c}}}{dt}+I\bm{\omega})=mR\bm{n}\times\\
 \bigg(\bm{g}-\bm{\Omega}\times(\bm{\Omega}\times \bm{r_{c}})-2\bm{\Omega}\times\frac{d\bm{r_{c}}}{dt} \bigg)+I\bm{\omega}\times\bm{\Omega}
\end{aligned}
\label{eq:EOM}
\end{equation}

To sum up, we have collected nine independent variables for our system: the position of the center of mass $(x_{c},y_{c},z_{c})$, the spin angular velocity of the ball $(\omega_{x},\omega_{y},\omega_{z})$, and the position of the contact point $(x_{o},y_{o},z_{o})$ contained in the expression of the normal vector $\bm{n}$. Noticing that the geometric constraint Eq.~\ref{eq:PositionConstraion} is a holonomic constraint with no velocity--dependence, we are able to eliminate the coordinates $(x_{c},y_{c},z_{c})$, and therefore the total degrees of freedom of the system are reduced to six.

For the pure--rolling case, the kinematic constraint Eq.~\ref{eq:PureRolling} is a non--holonomic constraint with a velocity--dependent term $d\bm{r_{c}}/dt$.\cite{borisov2002rolling, flannery2011elusive} Since we cannot further eliminate more independent variables with this non--integrable constraint, we have to simultaneously solve the rest two differential vector equations (the constraint Eq.~\ref{eq:PureRolling} and the governing Eq.~\ref{eq:EOM}) to find the evolution of the rest six coordinates $(x_{o},y_{o},z_{o})$ and $(\omega_{x},\omega_{y},\omega_{z})$. An example of the trajectory of the point $O$ of a pure--rolling ball is presented in Fig.~\ref{fig:traj}(a).

For the frictionless--slipping case, since Eq.~\ref{eq:PureSlipping} is a simple holonomic constraint, the total degrees of freedom of the system are further reduced to three. Therefore, one can find the evolution of the system by solely solving the Eq.~\ref{eq:EOM}.

\section{Linearized Governing Equations}

The rigorous motion equations derived in the previous section is nonlinear due to the form of the normal vector $\bm{n}$. To analyze the stability of the system, we linearize the system by expanding the variables in the vicinity of the equilibrium point.

\begin{figure}[htbp]
\centering
\includegraphics[width=0.25 \textwidth]{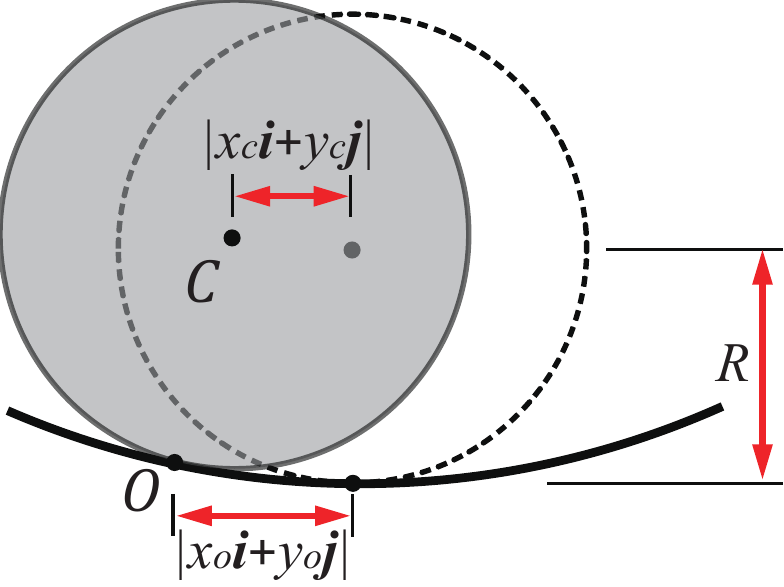}
\caption{Infinitesimal displacement of the rigid ball from its equilibrium position (dashed circle). The vertical displacement $z_{o}$ is one order smaller than the horizontal ones. The relation between the position vectors of the contact point $O$ and the center of mass $C$ is $\bm{r_{c}}=(1+\Lambda)\bm{r_{o}}+R\bm{k}$.}
\label{fig:InfinitesimalConfig}
\end{figure}

We assume that the horizontal displacement of the contact point at the equilibrium position is small compared to the size of the saddle, i.e. $x_{o}/a\sim y_{o}/a \sim O(\epsilon)$. On making use of the equation of the saddle--surface (Eq.~\ref{eq:SaddleSurface}), it is easy to find that the vertical displacement $z_{o}/a\sim O(\epsilon^{2})$, so vertical motion can be neglected. Hence in the following discussions, by collecting all the dimensionless quantities to the order of $\epsilon$, we can neglect the $z$ component of $\bm{r_{o}}$ and linearize it to be $\bm{r_{o}}\approx x_{o}\bm{i}+y_{o}\bm{j}$ (see Fig.~\ref{fig:InfinitesimalConfig}). In addition, the expression of the unit vector $\bm{n}$ can be linearized to be
\begin{equation}
\bm{n}=\frac{\Lambda}{R}\bm{r_{o}}+\bm{k}
\label{eq:LinearizedNormalVector}
\end{equation}
where $\Lambda$ is a matrix defined as
\begin{equation}
\Lambda=
\left(
  \begin{array}{ccc}
    -2R/a & 0   & 0\\
    0    & 2R/a & 0\\
    0    & 0   & 0\\
  \end{array}
\right)
\label{eq:CoordinateTransform}
\end{equation}
The position vector of the center of mass $\bm{r_{c}}$, together with its $n$th derivatives, can also be expressed in terms of $\bm{r_{o}}$ according to Eq.~\ref{eq:PositionConstraion}
\begin{equation}
\left\{
\begin{aligned}
&\bm{r_{c}}=(1+\Lambda)\bm{r_{o}}+R\bm{k}\\
&\frac{d^{n}}{dt^{n}}\bm{r_{c}}=(1+\Lambda)\frac{d^{n}}{dt^{n}}\bm{r_{o}}\qquad (n\geq 1)\\
\end{aligned}
\right.
\label{eq:LinearizedPositionConstrain}
\end{equation}
We further define a matrix $\tilde{\Lambda}$ as follows
\begin{equation}
\tilde{\Lambda}=\Lambda(1+\Lambda)^{-1}=
\left(
  \begin{array}{ccc}
    -2R/(a-2R) & 0       & 0\\
    0       & 2R/(a+2R) & 0\\
    0       & 0       & 0\\
  \end{array}
\right)
\label{eq:CoordinateTransformTilde}
\end{equation}
Finally, by substituting these linear expressions (Eq.~\ref{eq:LinearizedNormalVector} and Eq.~\ref{eq:LinearizedPositionConstrain}) into the rigorous equation of motion Eq.~\ref{eq:EOM}, collecting all terms to $O(\epsilon)$ and using the identity $\bm{k}\times (\bm{v}\times \bm{k}) \equiv\bm{v}$ for an arbitrary vector $\bm{v}$ perpendicular to $\bm{k}$, the linearized equation of motion for the pure--rolling case turns out to be
\begin{equation}
\begin{aligned}
(1+\alpha)\frac{d^{2}\bm{r_{c}}}{dt^{2}}+(2+\alpha)\bm{\Omega}\times\frac{d\bm{r_{c}}}{dt}-(\frac{g}{R}\tilde{\Lambda}+\bm{\Omega}^{2})\bm{r_{c}}\\
+\alpha \tilde{\Lambda}\Big(\frac{d}{dt}+\bm{\Omega}\times\Big)\bm{r_{c}}\times \bm{\omega}=0
\end{aligned}
\label{eq:LinearizedEOMforPureRolling}
\end{equation}
and that for the frictionless--slipping case is
\begin{equation}
\frac{d^{2}\bm{r_{c}}}{dt^{2}}+2\bm{\Omega}\times\frac{d\bm{r_{c}}}{dt}-(\frac{g}{R}\tilde{\Lambda}+\bm{\Omega}^{2})\bm{r_{c}}-\alpha (\bm{\omega}\times\bm{\Omega})\times\bm{k}=0
\label{eq:LinearizedEOMforPureSlipping}
\end{equation}

One can find that the spin angular velocity $\omega_z$ is constant in both the pure--rolling and the frictionless--slipping cases under linear approximation. For the pure--rolling case, from Eq.~\ref{eq:PureRolling}, we know that $\omega_x, \omega_y$ is small. In Eq.\ref{eq:LinearizedEOMforPureRolling}, they are further multiplied by $r_{cz}$, thus the effect of the horizontal angular velocity of the ball in this case can be neglected. Whereas in the frictionless--slipping case, $\omega_x, \omega_y$ are constant. However, non-zero $\omega_x, \omega_y$ will induce a constant force on the ball, then it cannot be stabilized near the origin of the saddle. So we consider $\bm{\omega}\approx \omega_z\bm{k}$ for both cases.

Notably, an operator $(d/dt+\bm{\Omega\times})$ shows up in Eq.~\ref{eq:LinearizedEOMforPureRolling}, which reminds us of the transformation of the differential operator $(d/dt)$ in the laboratory--frame form to its rotating--frame form\cite{lu1999university, goldstein1965classical, jackson2015hurricane}
\begin{equation}
\bigg(\frac{d}{dt}\bigg)_{Lab}=\bigg(\frac{d}{dt}\bigg)_{Rot}+\bm{\Omega}\times
\label{eq:DifferentialTransform}
\end{equation}
On making use of this transformation, we express the last driving term on the left--hand--side of Eq.~\ref{eq:LinearizedEOMforPureRolling} as $\alpha\tilde{\Lambda}(\bm{v_{Lab}}\times\bm{\omega})$, which turns out to be a typical Lorentz--like force. For a charged particle $e$ with mass $m$ and velocity $\bm{v}$ in the laboratory frame, it `feels' a Lorentz force $e\bm{v}\times\bm{B}$ in an external magnetic field $\bm{B}$. Likely in our mechanical model, given a specific vertical spin--angular velocity $\bm{\omega}$, the ball would `feel' a side force whose direction is perpendicular to its velocity in the laboratory frame as if a charged particle `feels' a Lorentz force in a magnetic field.

On a more physical level, this intriguing Lorentz--like driving term appears due to the fact that the spin--orbit interaction couples the spin of the pure--rolling ball to its external orbital motion around the equilibrium point. In Section.~\ref{sec:emanalogy}, we will further discuss how this Lorentz--like term can mimic the magnetic field for a specific quadrupole ion trap.
\begin{figure}[htbp]
\centering
\includegraphics[width=0.43 \textwidth]{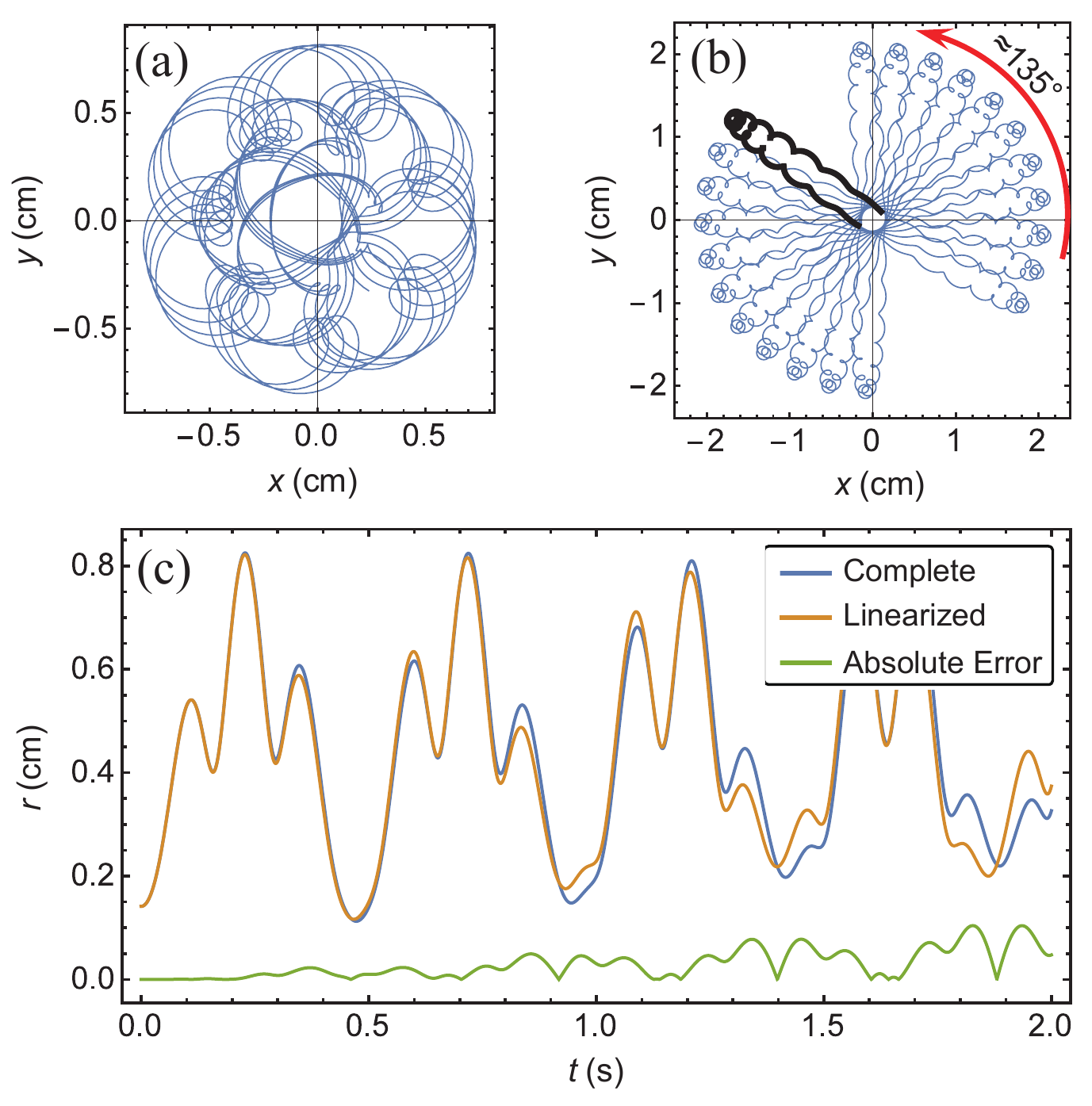}
\caption{Numerical trajectories of the contact point $O$ of (a) a pure--rolling ball with the radius of $2.0 cm$ and (b) a frictionless--slipping ball with radius $R\rightarrow0$ initially released at $x=y=0.1 cm$ with zero spin angular velocity in a rotating saddle with $a=0.159 m$ and $\Omega=30 rad/s$. The total evolving time of (a) and (b) are $10.0 s$ and $33.5 s$, respectively. A single petal of trajectory of the frictionless--slipping case is marked in thick black line. It can be estimated that the angle of precession of trajectory (b) agrees well with the theoretical value of $135^{\circ}$. (c) Comparison between the numerical results of $r_o=\sqrt{x_o^2+y_o^2}$ calculated from the rigorous equation and the linearized equation for the rigid ball.}
\label{fig:traj}
\end{figure}

The governing vector equations for the linearised system (Eq.~\ref{eq:LinearizedEOMforPureRolling} and Eq.~\ref{eq:LinearizedEOMforPureSlipping}) can be consistently separated into the following form:
\begin{equation}
\left\{
\begin{aligned}
\ddot{x}_{c}=Ax_{c}+B\dot{y}_{c}\\
\ddot{y}_{c}=Cy_{c}+D\dot{x}_{c}
\end{aligned}
\right.
\label{eq:LinearizedEOM}
\end{equation}
where the $ABCD$ coefficients for the pure--rolling and the frictionless slipping cases are shown in Table~\ref{tab:abcdCoefficients}. The correctness of these linearized equations is simply verified in Fig.~\ref{fig:traj}(c) by numerically solving the trajectory of a pure--rolling ball and comparing the result with the rigorous one.

Interestingly, a large discrepancy in trajectories is found between the pure--rolling case and the frictionless--slipping case (see Fig.~\ref{fig:traj}). For the frictionless--slipping case (which is the same as the mass--point model when $R\rightarrow 0$), a regular precession is observed. The angle of precession within an interval of $33.5s$ agrees very well with the value predicted by the mass--point model\cite{kirillov2016rotating} (see Fig.~\ref{fig:traj}(b))
\begin{equation}\label{}
\Theta_{p}=\Omega_{p}t=\frac{g^{2}}{2a^{2}\Omega^{3}}t\cong 135.03^{\circ}
\end{equation}
Whereas for the pure-rolling case, the ball whirls about the origin in a different manner (see Fig.~\ref{fig:traj}(a)). The trajectories of these two cases do not coincide even under the limit when $R/a \rightarrow 0$ (also see Eq.~\ref{eq:LinearizedEOM} and Table.~\ref{tab:abcdCoefficients} for comparison). This difference in orbital patterns can be explained by analyzing the eigen--frequencies of the system. For the frictionless--slipping case without spin--orbit interaction, the magnitudes of the two pairs of the eigen--frequencies are close to each other when $\Omega^2\gg g/a$, forming a precessing secular motion and a micro--oscillation. For the pure--rolling case with the strongest spin--orbit interaction, however, the eigen--frequencies are significantly different, making the trajectory more complex.

\section{Critical Angular Velocity}

\begin{table}
\caption{\label{tab:abcdCoefficients}The four coefficients for the linearized governing equations under the pure--rolling and the frictionless--slipping conditions.}
\begin{ruledtabular}
\renewcommand\arraystretch{2.5}
\begin{tabular}{ccc}
     & \textbf{Pure--rolling}                                                        &\textbf{Frictionless--slipping}\\
\itshape $A$            &$\displaystyle\frac{(a-2R)\Omega^{2}+2\alpha R \omega_{z}\Omega-2g}{(1+\alpha)(a-2R)}$    &$\displaystyle\frac{(a-2R)\Omega^{2}-2g}{a-2R}$\\
\itshape $B$            &$\displaystyle\frac{(2+\alpha)(a+2R)\Omega-2\alpha R \omega_{z}}{(1+\alpha)(a+2R)}$         &$\displaystyle2\Omega$\\
\itshape $C$            &$\displaystyle\frac{(a+2R)\Omega^{2}-2\alpha R \omega_{z} \Omega+2g}{(1+\alpha)(a+2R)}$     &$\displaystyle\frac{(a+2R)\Omega^{2}+2g}{a+2R}$\\
\itshape $D$            &$\displaystyle\frac{-(2+\alpha)(a-2R)\Omega-2\alpha R \omega_{z}}{(1+\alpha)(a-2R)}$        &$\displaystyle-2\Omega$\\
\end{tabular}
\end{ruledtabular}
\end{table}

Based on Eqs.~\ref{eq:LinearizedEOM}, we further analyze the stability of the rigid ball near the equilibrium point in this section. Consider now a trial solution of the form
\begin{equation}
x_{c}=X_{0}e^{i\varpi t},\qquad y_{c}=Y_{0}e^{i\varpi t}
\label{eq:TrialSolution}
\end{equation}
where $\varpi$ is the eigen circular frequency of the system. In general, $\varpi$ is a complex number: its real part $Re(\varpi)$ determines the frequency of oscillation, and its imaginary part $Im(\varpi)$ determines the stability behaviour near the equilibrium point. In the case of a positive imaginary part, the solution decays over time, and the system is stable, whereas in the case of a negative imaginary part, the solution exponentially grows over time and thus the system becomes unstable.
\begin{figure}[htbp]
\centering
\includegraphics[width=0.40 \textwidth]{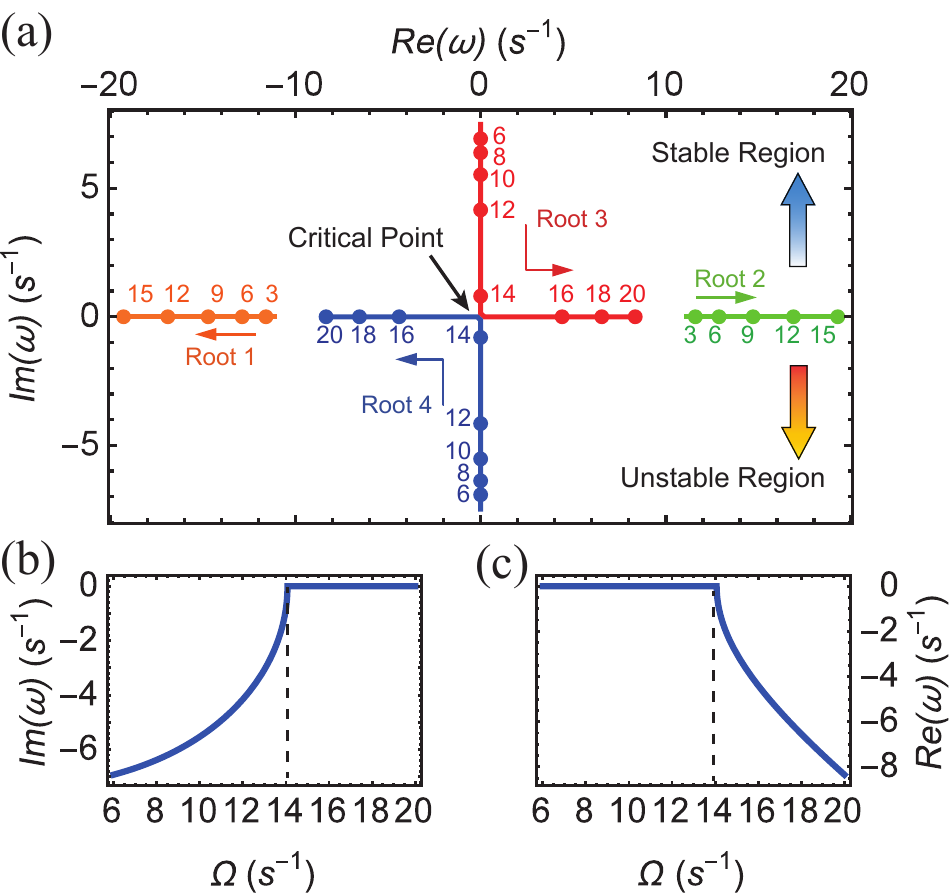}
\caption{(a) Four loci of eigen frequencies $\varpi$ on the complex plane for the system with ball radius $R=0.03m$ and saddle parameter $a=0.159m$ under the pure--rolling condition. The numbers marked beside each point denote the corresponding angular velocities $\Omega$ of the rotating saddle. (b)(c) The imaginary and the real parts of the Root 4 as a function of $\Omega$. The critical angular velocity $\Omega_{crt}$ (beyond which all the eigen frequencies, including the Root 4, possess non--negative imaginary parts) is $14.07 s^{-1}$ according to Eq.~\ref{Eq:RigidBodyThreshold}.}
\label{fig:ComplexFrequency}
\end{figure}

\begin{figure*}[htbp]
\centering
\includegraphics[width=0.75 \textwidth]{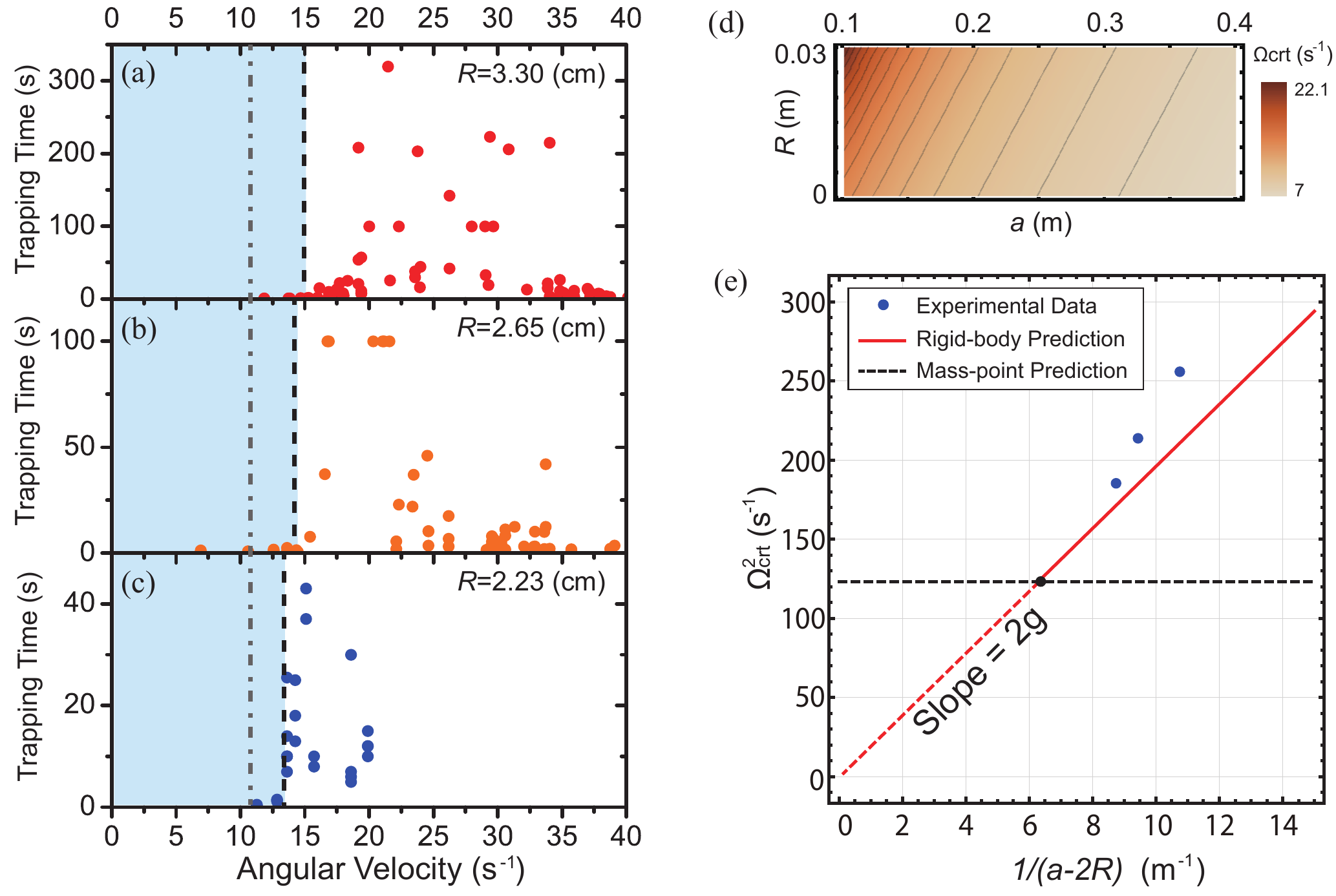}
\caption{The trapping time of rigid balls with radii of (a) $3.30cm$, (b) $2.65cm$ and (c) $2.23cm$ under different rotating speeds. The dashed lines denote the theoretical critical angular velocity $\Omega_{crt}$ predicted by the rigid--body model in Eq.~\ref{Eq:RigidBodyThreshold}, and the dot--dashed lines denote the theoretical $\Omega_{crt}$ predicted by the mass--point model in Eq.~\ref{eq:MassPointThreshold}. (d) The critical angular velocity $\Omega_{crt}$ for stabilizing rigid balls as a function of the saddle parameter $a$ and the ball radius $R$. (e) Comparisons among the critical angular velocities provided by the experimental data, the rigid--body model and the mass--point model.}
\label{fig:result}
\end{figure*}

Substituting Eq.~\ref{eq:TrialSolution} into Eq.~\ref{eq:LinearizedEOM} leads to the following algebraic equations
\begin{equation}
\left\{
\begin{aligned}
(A+\varpi^{2})\cdot X_{0}+i\varpi B\cdot Y_{0}&=0\\
i\varpi D\cdot X_{0}+(C+\varpi^{2})\cdot Y_{0}&=0
\end{aligned}
\right.
\label{eq:AlgebraicEOM}
\end{equation}
For non-trivial solutions, the coefficient determinant of the above algebraic equations must vanish, yielding the following eigen equation for the complex frequency $\varpi$:
\begin{equation}
\begin{aligned}
\left| \begin{array}{cc}
A+\varpi^{2} & i\varpi B\\
i\varpi D & C+\varpi^{2}\\
\end{array} \right|
&=\varpi^{4}+(A+C+BD)\varpi^{2}+AC\\
&=0
\end{aligned}
\label{eq:Determinant}
\end{equation}
For a rotating saddle system with given geometrical parameters, vertical spin angular velocity $\omega_{z}$ and a particular rotating speed $\Omega$, the four coefficients are fixed, and therefore yield four complex eigen frequencies. We plot the loci of the four eigen frequencies on the complex plane as parametric
trajectories of the rotation speed $\Omega$. As an example illustrated in Fig.~\ref{fig:ComplexFrequency}, when the rotating speed of the saddle exceeds a certain value $\Omega_{crt}$ (in this case greater than roughly $14s^{-1}$), all of these complex frequencies have non--negative imaginary parts, and the system is stabilized in the vicinity of the equilibrium point.

Such a critical rotating speed $\Omega_{crt}$ can even be derived analytically as follows. As mentioned above, the stability condition requires the imaginary parts of all of its four roots to be non--negative, leading to
\begin{equation}
\left\{
\begin{aligned}
&(A+C+BD)^{2}>4AC\\
&AC>0\\
&A+C+BD<0
\end{aligned}
\right.
\label{eq:Inequalities}
\end{equation}
In addition, since the ball is not in ideal one--point contact with the saddle surface in real circumstances, the spin of the ball will gradually be adjusted to $\omega_{z}\approx 0$ due to rotational friction ($\omega_{z}=\Omega$ in the laboratory frame) when moving in the vicinity of the equilibrium point (see a illustrational video at [URL will be inserted by AIP]). Also the radius of the ball $R$ cannot exceed $a/2$, otherwise it will be stuck on the surface. Using these two conditions associated with inequalities~\ref{eq:Inequalities}, the critical angular velocity $\Omega_{crt}$ for the rotating--saddle to trap a rigid ball with radius $R$ turns out to be
\begin{equation}
\Omega \geq \Omega_{crt} = \sqrt{\frac{2g}{a-2R}}
\label{Eq:RigidBodyThreshold}
\end{equation}
Using this formula we find that the critical rotating speed beyond which the system illustrated in Fig.~\ref{fig:ComplexFrequency} becomes stable is $14.07 s^{-1}$.

The fact that $\Omega_{crt}$ is identical for both the pure--rolling and the frictionless--slipping cases brings about convenience for our experiments, in that we cannot actually control the ball to be rolling or slipping in real circumstances. In addition, since $\Omega_{crt}$ is independent of coefficient $\alpha$, the radial mass distribution (i.e. whether the ball is solid or hollow) does not influence the stability of the system.

To verify the critical angular velocity for confining rigid balls, we fabricated a saddle--shaped surface with a 3D printer. The geometrical parameter $a$ of the saddle was set to be $0.159m$. The saddle was driven by a $24V$ DC motor to rotate around the vertical axis. By adjusting voltage on the DC motor between $0V$ and $24V$, we were able to control the rotating speed $\Omega$ of the saddle, which was further measured by a laser tachometer.

We used polyfoam balls for our experiment. The advantage of using polyfoam balls is that their small weights would have little influence on the rotating speed $\Omega$ of the saddle. By carefully placing polyfoam balls with different sizes onto the center of the rotating saddle and recording their motions with a high--speed camera, we can measure the time the ball is trapped by the saddle. This procedure was repeated for a number of times for the balls with radii of $3.30cm$, $2.65cm$ and $2.23cm$. The results are shown in Fig.~\ref{fig:result}(a), (b) and (c).

As can be seen in Fig.~\ref{fig:result}(a)(b)(c), the trapping time of the polyfoam balls dramatically increases after certain thresholds of rotating speed $\Omega$. These thresholds lies closer to the critical angular velocity $\Omega_{crt}$ predicted by our rigid--body model (dashed line) than that predicted by the mass--point model (dot--dashed line). Such a discrepancy between the rigid--body model and the mass--point model becomes considerably larger as the radius of the ball increases (see Fig.~\ref{fig:result}(d)).

To better evaluate the rationality of our model, we plot $\Omega_{crt}^{2}$ versus $1/(a-2R)$, because according to Eq.~\ref{Eq:RigidBodyThreshold}, $\Omega_{crt}^{2}$ should be proportional to $1/(a-2R)$ with the slope of $2g$.
The experimental critical angular velocities are chosen to be those velocities for which the trapping time of the polyfoam ball first exceeds
$60\pi/\Omega$, roughly 10 seconds in our cases.

The final results are presented in Fig.~\ref{fig:result}(e). Compared with the mass--point prediction, much better agreements can be found between the experimental data and the theoretical prediction of rigid--body model.

\section{High-speed Instability}
\begin{figure}[htbp]
\centering
\includegraphics[width=0.40 \textwidth]{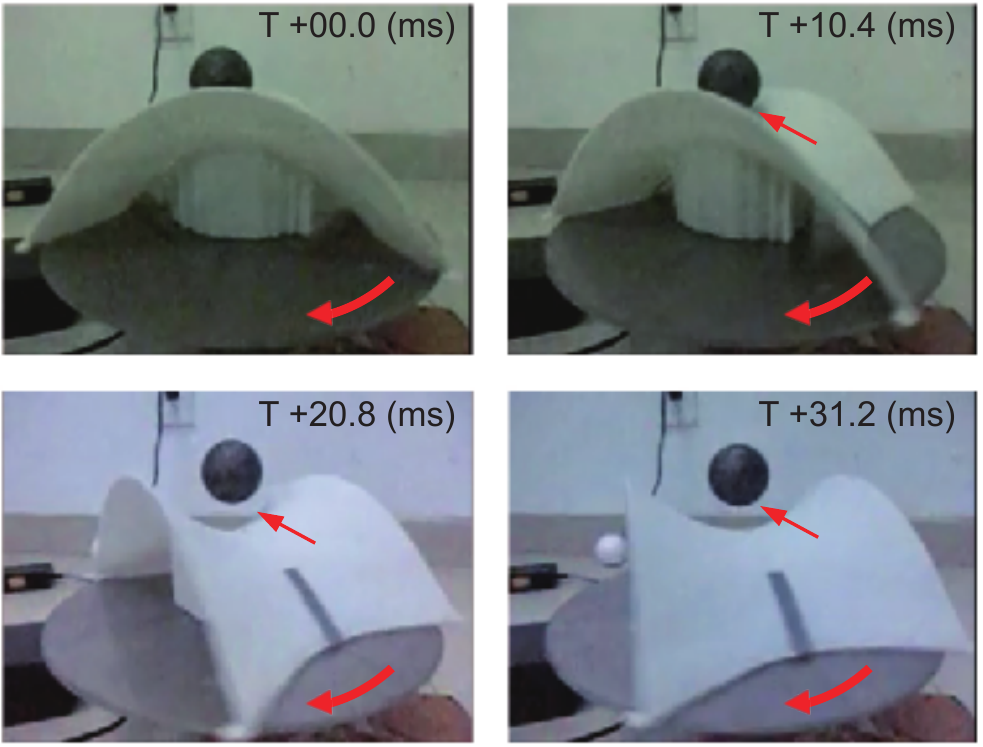}
\caption{Four consecutive snapshots of a polyfoam ball with radius of $2.65cm$ detaching from the saddle surface (marked by red arrows). The rotating speed $\Omega$ of the saddle is $43.6 s^{-1}$, which is about three times greater than the critical angular velocity $12.2 s^{-1}$.}
\label{fig:detach}
\end{figure}

By analyzing the eigen frequencies of the linearized system, it has been noticed that a mass--point constrained on non--symmetric rotating saddles (i.e. saddles possessing different radii of curvature in two directions) would again lose stability as the rotating speed $\Omega$ exceeds an upper bound.\cite{kirillov2013exceptional} For symmetric saddle, however, either the mass--point model or the rigid--body model till now are only able to predict a lower bound for stabilizing the balls. But we do observe that the trapping time of a polyfoam ball becomes considerably shorter as the rotating speed of our symmetric saddle becomes high (see Fig.~\ref{fig:result}(a)(b)(c))---there seems to be a ``vague'' upper limit of $\Omega$ to confine the ball in our saddle.

A typical example of such high--speed instability is recorded by a high--speed camera in Fig.~\ref{fig:detach} (see a high-speed video at [URL will be inserted by AIP]). As can be seen, as the rotating speed $\Omega$ reaches $43.6 s^{-1}$ (which is about three times greater than the critical rotating speed $12.2 s^{-1}$), the ball losses contact with the saddle surface and jumps off.

Here we come up with an important mechanism that can lead to such high--speed instability. Glancing back over the dynamic model we have built, we find the interaction between the saddle and the rigid ball does not allow the supporting force $\mathbf{F_{n}}$ to point in toward the surface. This means that $\mathbf{F_{n}}\cdot\mathbf{n}<0$ is forbidden in real circumstances: as the supporting force becomes less than zero, the ball simply loses contact with and jumps off the saddle surface.

\begin{figure}[htbp]
\centering
\includegraphics[width=0.45 \textwidth]{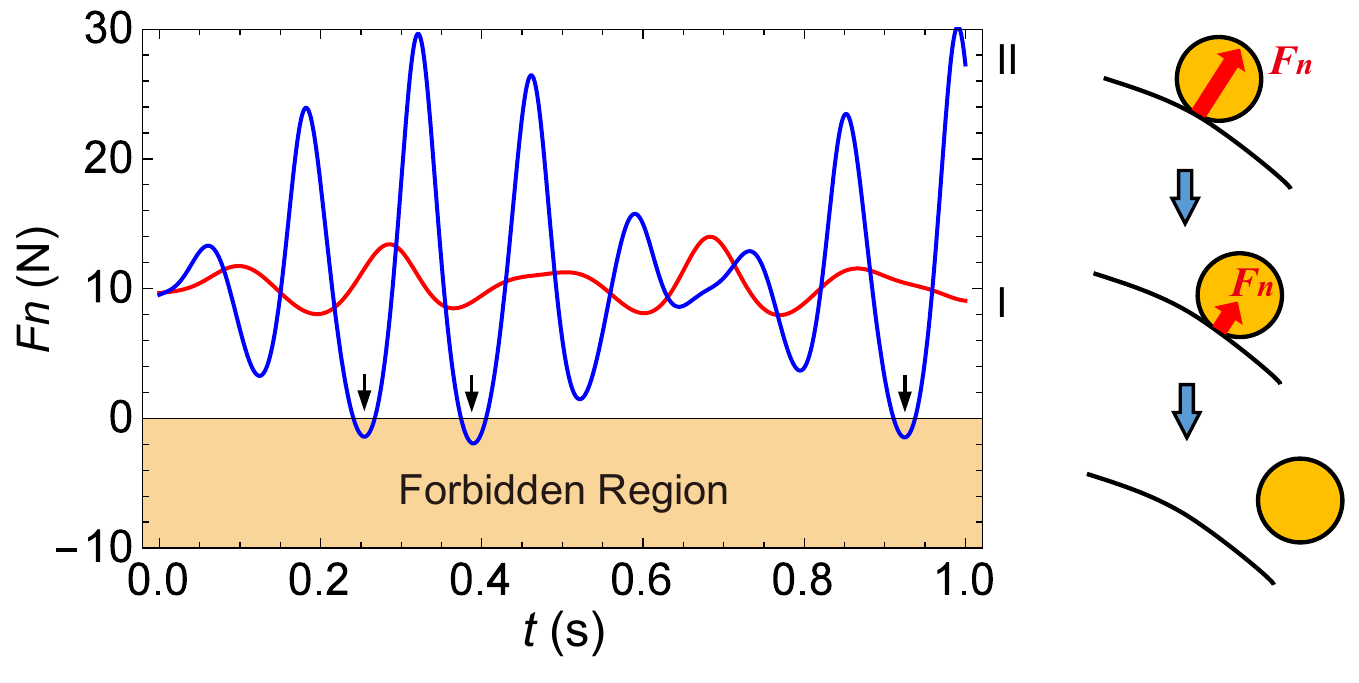}
\caption{Theoretical supporting forces provided by the saddle with geometric parameter $a=0.159m$. The ball with radius $3.30 cm$ is initially placed at the position $x=0.5cm$ and $y=0.5cm$ at $t=0$. The rotating speed $\Omega$ for the red line (I) and the blue line (II) is $20 s^{-1}$ and $30 s^{-1}$ respectively. Negative supporting forces (marked by black arrows) are forbidden, which, in real circumstances, are conditions that the ball detaches from the saddle surface.}
\label{fig:support}
\end{figure}

\begin{figure}[htbp]
\centering
\includegraphics[width=0.48 \textwidth]{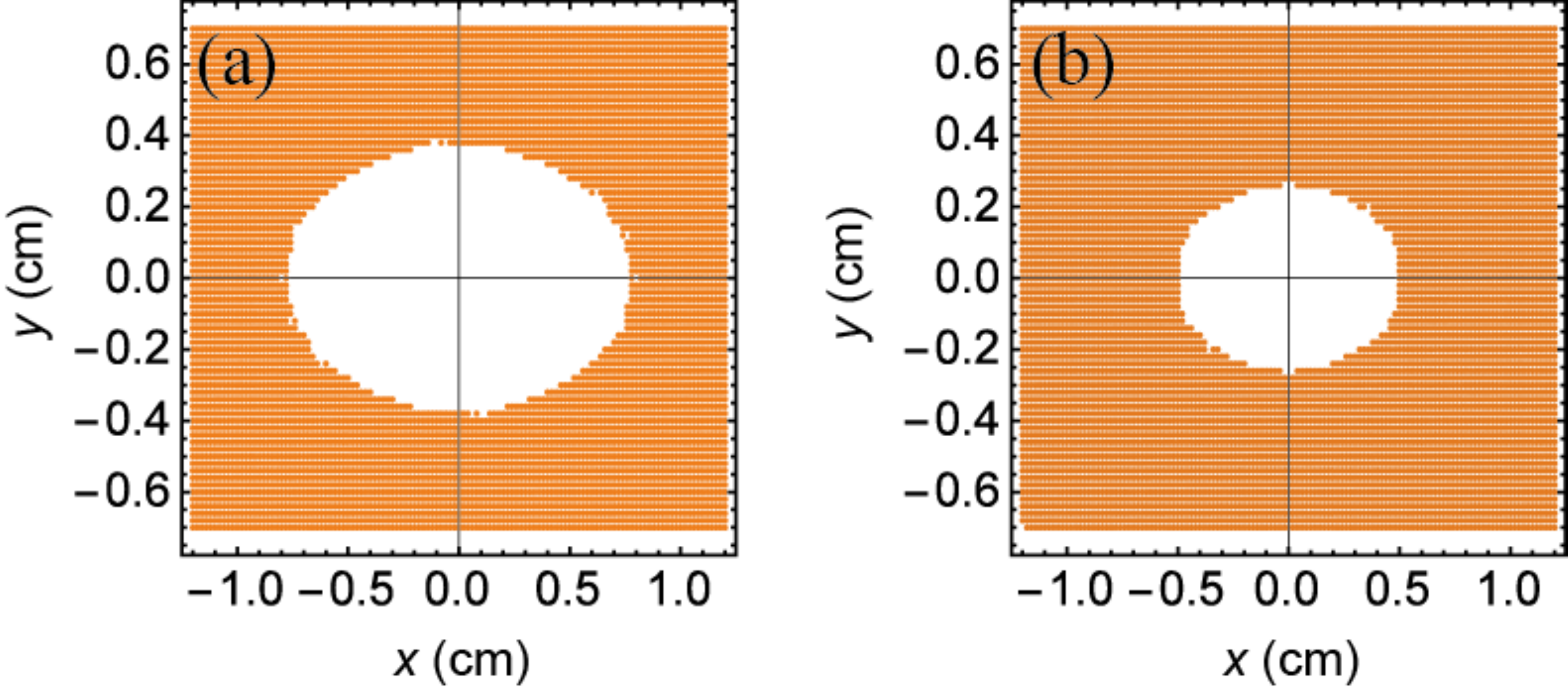}
\caption{The white region indicates the feasible initial positions for (a) $\Omega=35s^{-1}$ and (b) $\Omega=45s^{-1}$ where a ball can be placed at without jumping off the saddle during its subsequent evolution. The area gets noticeably smaller as $\Omega$ increases.}
\label{fig:feasible}
\end{figure}

As an example, we track the supporting force provided by the saddle surface as a function of time (See Fig.~\ref{fig:support}). We calibrate the rotating speed $\Omega$ to be $20 s^{-1}$ and $30 s^{-1}$ which are all beyond $\Omega_{crt}$. When $\Omega$ is $20 s^{-1}$, the supporting force remains positive over time (see red line in Fig.~\ref{fig:support}). But when $\Omega$ reaches $30 s^{-1}$, we find a negative supporting force appears, which is actually forbidden in real cases (a surface cannot ``pull'' an object to itself). The existence of this forbidden interaction is why high--speed instability occurs.

\begin{figure*}[htbp]
\centering
\includegraphics[width=0.72 \textwidth]{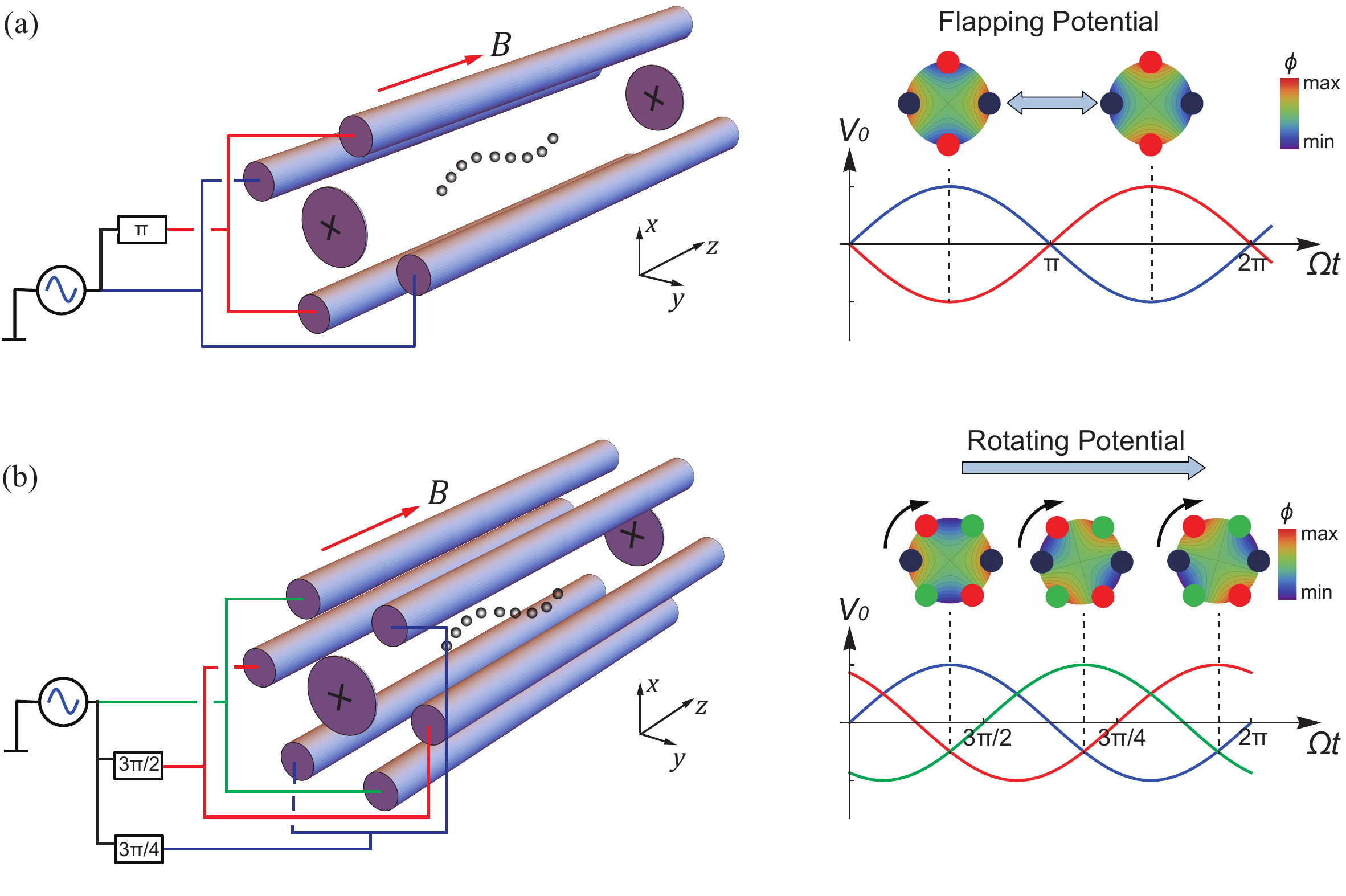}
\caption{Configurations of typical quadrupole ion traps consisting of (a) four rod electrodes (a.k.a. the Paul trap) and (b) six rod electrodes (a.k.a the rotating--radio--frequency (or RRF) trap). A string of ions is confined inside each trap. The Paul trap uses an oscillating or ``flapping'' quadrupole potential, whereas the RRF trap uses a rotating quadrupole potential to stabilize the ions in the vicinity of the central line.}
\label{fig:QITConfig}
\end{figure*}

The condition that yields high--speed instability is not only dependent on the geometric and kinematic parameters of the system, but also on the initial conditions. For different rotating speed $\Omega$ of the saddle, we plot the feasible regions of the initial positions of the ball, within which the supporting force remains positive when evolving over time (see Fig.~\ref{fig:feasible})). As $\Omega$ becomes larger, the feasible region becomes smaller, and it becomes less likely to put the ball within the feasible region in real experiments for it to be trapped. Other systematic influences such as air flow and some defect of the apparatus may also contribute to the instability of the ball in real experiments.

\section{Electromagnetic Analogy}\label{sec:emanalogy}

The analogy with a ball rolling on a rotating saddle provides an intuitive way to interpret the mechanism of confining ions with a type of quadrupole ion trap called the Paul trap. However, it has been pointed out by several papers that this analogy, to some extent, is quantitatively inaccurate on close scrutiny.\cite{hasegawa2005rotating, thompson2002rotating, Shapiro2001Rotating} In this section, we are going to review these quantitative analyses, and compare them with the rigid--body model of rotating saddle trap to find a more accurate mechanical correspondence.

Typically, quadrupole ion traps are composed of four or six rod electrodes, two end--cap electrodes and sometimes an external magnetic field (see Fig.~\ref{fig:QITConfig}). The quadrupole trap with four rod electrodes is also known as the Paul trap.\cite{paul1990electromagnetic} Each rod electrode of the Paul trap is connected to its diagonally opposite one, and a sinusoidal voltage with amplitude $V_{0}$ is applied between these two electrode pairs. With this configuration, a phase difference of $\pi$ is induced between the voltages on the electrode pairs, and an oscillating or ``flapping'' saddle--like quadrupole electric potential is generated in the vicinity of the central line on $XOY$ cross--section\cite{foot2004atomic} whose expression is
\begin{equation}
\phi=\phi_{0}+\frac{V_{0}}{2r_{0}^{2}}\cos(\Omega t)(x^{2}-y^{2})
\label{eq:EPotentialPaulTrap}
\end{equation}
where $2r_{0}$ is the distance between two diagonal electrodes and $\Omega$ is the circular frequency of the applied AC voltage. Although the Paul trap utilizes the same concept of stabilizing the equilibrium by varying the potential over time\cite{kirillov2016rotating} as the rotating saddle, it was noticed\cite{hasegawa2005rotating, thompson2002rotating} that the ``flapping'' quadrupole potential is not quantitatively the same as the rotating-saddle trap.

Another kind of quadrupole ion traps composed of six rode electrodes is known as the rotating radio--frequency (RRF) trap (see Fig.~\ref{fig:QITConfig}(b)).\cite{hasegawa2005rotating, huang1997steady, huang1998precise, hasegawa2005stability} In this configuration, they utilized a three--phase AC source which is often used to generate a rotating electromagnetic field\footnote{A very comprehensive introduction on this topic can be found on Wikipedia \url{https://en.wikipedia.org/wiki/Rotating_magnetic_field}.}. Each pair of rods are connected to an output wire of a three--phase AC source, and thus the quadrupole electric potential generated in the vicinity of the central line has the form of
\begin{equation}
\phi=\phi_{0}+\frac{V_{0}}{2r_{0}^{2}}(x'^{2}-y'^{2})
\label{eq:EPotentialRRFTrap}
\end{equation}
where $x'=x \cos(\Omega t)-y \sin(\Omega t)$ and $y'=x \sin(\Omega t)+y \cos(\Omega t)$ are coordinates in the rotating frame. The rotating nature of the quadrupole field produced by RRF trap thus qualitatively mimics the gravitational potential in a rotating--saddle trap $mg(x'^{2}-y'^{2})/a$.

In addition to the rotating quadrupole field, the RRF trap contains also a constant magnetic field $\bm{B}$ that further modulates the oscillating frequency of the plasma confined inside the trap. What is interesting is that the Lorentz force $e\bm{v}\times \bm{B}$ on every single ion in the RRF trap, as mentioned previously, is of the same form of the driving term $\alpha \tilde{\Lambda} (\bm{v_{Lab}}\times\bm{\omega})$ appeared in Eq.~\ref{eq:LinearizedEOMforPureRolling} for a pure--rolling rigid ball in a rotating saddle. This indicates that the spin angular velocity $\bm{\omega}$ of a pure--rolling ball is an analogy to the constant magnetic field $\bm{B}$ in the RRF traps.

Remarkably, the motion equation of ions in RRF traps\cite{hasegawa2005stability} is the same as Eq.~\ref{eq:LinearizedEOM}, with coefficients ABCD containing terms similar to those in Table.~\ref{tab:abcdCoefficients}:
\begin{equation}
\left\{
\begin{aligned}
&A=\Omega^2-\Omega\omega_B-k_{1} V_{0}+k_{2} V_{t},\quad B=2\Omega-\omega_B\\
&C=\Omega^2-\Omega\omega_B+k_{1} V_{0}+k_{2} V_{t},\quad D=-2\Omega+\omega_B
\end{aligned}
\right.
\label{comp2}
\end{equation}
where $V_{t}$ is the voltage of a static electric field for $z$ direction confinement, $\omega_B=eB/m$ is the cyclotron frequency of ions in the magnetic field and $k_{1}$, $k_{2}$ are two geometrical factors that are determined by the electrode configuration.

By scaling the four coefficients of the rotating saddle trap by a factor $(1+\alpha)$ and defining $p=\alpha \widetilde{\Lambda}_{11}, q=\alpha \widetilde{\Lambda}_{22}$, we obtain
\begin{equation}
\left\{
\begin{aligned}
&A=\Omega^{2}-p\Omega \omega_z-\frac{p}{R}g,\quad B=(2+\alpha)\Omega-q\omega_z\\
&C=\Omega^{2}-q\Omega \omega_z+\frac{q}{R}g,\quad D=-(2+\alpha)\Omega+p\omega_z
\end{aligned}
\right.
\label{comp3}
\end{equation}
Despite some geometric factors brought in either by the electrode configuration or by the geometry of the ball, the field $V_0$ acts just as the gravitational force $g$, and the spin of the ball along $z$ direction $\omega_z$ mimics the cyclotron frequency of charges in the magnetic field $B$ ($\omega_B=eB/m$). Therefore we conclude that the motion of a pure--rolling ball in the rotating saddle can quantitatively mimic the behavior of the ions confined in the RRF trap with a constant magnetic field.

\section{Conclusion}

We built a rigid--body mechanical model for the rotating--saddle traps. We would like to end our article by reviewing the similarities and differences between our rigid--body model and the mass--point model.

A modified critical angular velocity for the saddle to stabilize rigid balls is derived, taking the size of the ball into consideration (see Eq.~\ref{Eq:RigidBodyThreshold}). This critical angular velocity is identical for both the pure--rolling and the frictionless--slipping balls, and tends to that predicted by the mass--point model when $R\rightarrow 0$.

We found that the mass--point model is a limitation of the frictionless--slipping model when the radius of the ball $R\rightarrow 0$. They not only have the same orbital patterns (see Fig.~\ref{fig:traj}(b)), but also present exactly the same precessional behavior. On the contrary, the orbital pattern of a pure--rolling ball, due to the spin--orbit interaction, is different from the above two cases, even when the radius of the ball is negligible.

Finally, we figured out that our rigid body model is a mechanical analogy to the motion of ions in an RRF trap with magnetic field. Under linear approximation, the motion of the ball in $z$ direction can be neglected, thus its constant spin angular velocity $\bm{\omega}=\omega_z\bm{k}$ fixed in the $z$ direction can is analogous to the cyclotron frequency of charged particles in the magnetic field.

\section{Acknowledgement}
This work is partly funded by the National Natural Science Foundation of China (NNSFC--J1310026).

\appendix
\section{Derivation of Eq.~\ref{eq:TorqueAM}}
\label{appendix:A}
We consider the rigid body to be composed of $N$ small segments labeled by $i=1,2,3,\cdots,N$. The total torque on the rigid body with respect to the contact point $O$ is
\begin{equation}
\bm{M_{o}}=\sum_{i=1}^{N}\bm{r_{oi}}\times(\bm{F_{i}}+\sum_{j=1}^{N}\bm{F_{ij}})
\end{equation}
where $\bm{r_{oi}}$ is the position vector defined by $(\bm{r_{i}}-\bm{r_{o}})$, $\bm{F_{i}}$ is the external force on each segment, and $\bm{F_{ij}}$ is the internal force between segment $i$ and $j$. According to Newton's Third Law, all internal forces cancel out upon summation. On further making use of Newton's Second Law $\bm{F_{i}}=m_{i}d^{2}\bm{r_{i}}/dt^{2}$, we have
\begin{equation}
\bm{M_{o}}=\sum_{i=1}^{N}\bm{r_{oi}}\times m_{i}\frac{d^{2}\bm{r_{i}}}{dt^{2}}
\end{equation}

By using the total derivative theorem and the definition of angular momentum, we can finally prove
\begin{equation}
\begin{aligned}
  \bm{M_{o}}&=\frac{d}{dt}(\sum_{i=1}^{N}\bm{r_{oi}}\times m_{i}\frac{d\bm{r_{i}}}{dt})-\sum_{i=1}^{N}\frac{d\bm{r_{oi}}}{dt}\times m_{i}\frac{d\bm{r_{i}}}{dt}\\
  &=\frac{d\bm{L_{o}}}{dt}-\sum_{i=1}^{N}(\frac{d\bm{r_{i}}}{dt}-\frac{d\bm{r_{o}}}{dt})\times m_{i}\frac{d\bm{r_{i}}}{dt}\\
  &=\frac{d\bm{L_{o}}}{dt}+\frac{d\bm{r_{o}}}{dt}\times m\bm{v_{c}}
  \end{aligned}
\end{equation}

\end{document}